\documentclass[twocolumn]{aastex631}

\graphicspath{{./}{figs/}}
\usepackage{units}
\usepackage{mathtools}

\newcommand\foe{$\unit[10^{51}]{erg}$}
\newcommand\tenfoe{$\unit[10^{52}]{erg}$}

\begin{document}

\title{Explosions in Roche-Lobe Distorted Stars: Relativistic Bullets in Binaries}  

\author[0000-0003-3356-880X]{Marcus DuPont}
\affiliation{Center for Cosmology and Particle Physics, Physics Department, New York University 
New York, NY, 10003, USA}
\affiliation{Max-Planck-Institut f{\"u}r Astrophysik,
Karl-Schwarzschild-Stra$\beta$e 1, D-85741 Garching, Germany}

\author[0000-0002-0106-9013]{Andrew MacFadyen}
\affiliation{Center for Cosmology and Particle Physics, Physics Department, New York University 
New York, NY, 10003, USA}

\author[0000-0001-9336-2825]{Selma de Mink}
\affiliation{Max-Planck-Institut f{\"u}r Astrophysik,
Karl-Schwarzschild-Stra$\beta$e 1, D-85741 Garching, Germany}
\affiliation{Astronomical Institute Anton Pannekoek, Amsterdam University, Science Park 904, 1098 XH, Amsterdam, Netherlands}



\begin{abstract}
State-of-the-art surveys reveal that most massive stars in the universe evolve in close binaries. Massive stars in such systems are expected to develop aspherical envelopes due to tidal interactions and/or rotational effects. Recently, it was shown that point explosions in oblate stars can produce relativistic equatorial ring-like outflows. Moreover, since stripped-envelope stars in binaries can expand enough to fill their Roche lobes anew, it is likely that these stars die with a greater degree of asphericity than the oblate spheroid geometry previously studied. We investigate the effects of this asymmetry by studying the gas dynamics of axisymmetric point explosions in stars in various stages of filling their Roche lobes. We find that point explosions in these pear-shaped stars produce trans-relativistic ejecta that coalesces into bullets pointed both toward and away from the binary companion. We present this result and comment on key morphological differences between core-collapse explosions in spherical versus distorted stars in binary systems, effects on gravitational wave sources, and observational signatures that could be used to glean these explosion geometries from current and future surveys.
\end{abstract}



\section{Introduction} \label{sec:intro}

  Massive stars predominantly form in binary and multiple systems and the majority of them will interact and exchange mass with at least one companion star \citep{Sana+2012, Moe+diStefano+2017}.   
Because these stars are so massive, they will die as core-collapse supernovae (CCSNe), explosions that synthesize many of the heavy elements and drive outflows from galaxies \citep[e.g.][]{Fowler+Hoyle-1964,Woosley-2002,Efstathiou2000}.


The shape of a star in a binary system can deviate significantly from spherical symmetry. This is particularly true when a star expands to fill its Roche lobe. This forces the star to become ``pear-shaped'' due to the combined effect of the gravitational pull of the companion and the centrifugal effect from the motion around the common center of mass. In this paper, we explore what happens if a supernova explosion occurs in such a highly asymmetric, Roche-lobe-deformed star.  

Explosions in non-spherical stars are of interest because of the effects that an oblique shock can have when breaking out of the stellar surface. The non-radial motions can dramatically alter the dynamics and emission of shock emergence \citep{Matzner+2013}.  In a previous paper, \citet{DuPont+2022} numerically explored the effect of point explosions in rapidly rotating, oblate progenitors and showed how these produce relativistic equatorial ring-like outflows.  Here we build upon this work and extend this to the pear-shaped stars in binary systems. 

The fraction of core-collapse supernovae progenitors that are significantly Roche-lobe deformed is not very well known, but binary evolutionary simulations suggest that this may be common for lower mass progenitors of stripped-envelope supernovae (type Ib/c and type IIb, \citealt{Filippenko-1997}). These are produced in binary systems where the progenitor loses most of its H-rich envelope during a first mass transfer phase  \citep{Kippenhahn+Weigert+1967, Gotberg+2017}. The resulting stripped star first contracts as a hot helium star. \citet{Drout+2023} and \citet{Gotberg+2023} recently presented the discovery and analysis of such stars, which so far had been missing. 

During their later evolution, helium stars in the mass range $2-4M_\odot$ can expand significantly to become helium giants \citep[e.g.,][]{Divine+1965,Habets+1986,Dewi+Pols+2002,Dewi+Pols+2003,Yoon+2010,Sravan+2019}.  If they retain a layer of hydrogen, the expansion is even more drastic \citep[][]{Laplace+2020, Klencki+2022}. This implies that these low-mass stripped progenitors are (close to) filling their Roche lobes at the onset of core collapse and therefore deformed. 

Of special interest are supernova explosions in Roche-lobe deformed stripped stars where the companion star is a neutron star. These systems may lead to ultra-stripped SNe \citep{Tauris+2015}.
This is the last stage in the pathway towards the formation of double neutron stars and gravitational wave sources \citep{Tauris+2017}. In this work, we present a suite of two-dimensional axisymmetric simulations of point explosions in a $3M_\odot$ pre-supernova helium star at various stages of filling its Roche lobe, described in Section~\ref{sec:setup}. We study how the shock propagates and breaks out from the surface. We report on the formation of a trans-relativistic ejecta pointed both away and towards the companion in Section \ref{sec:results}. We discuss the astrophysical implications and observational consequences in Section~\ref{sec:discussion}. Section~\ref{sec:summary} provides a summary with conclusions.


\section{Numerical Setup} \label{sec:setup}
\subsection{Governing Equations}
The governing hydrodynamical equations in the special relativistic regime describe the conservation of baryon number, 
\begin{equation}\label{eq:baryon}
    \partial_\mu (\rho u^\mu) = 0,
\end{equation}
and conservation of stress-energy,
\begin{equation}
    \partial_\nu T^{\mu \nu} = 0,
\end{equation}
where we work in the fluid frame so that $T^{\mu \nu} = \rho h u^\mu u^\nu + p\eta^{\mu \nu}$. In these equations,
$h = 1 + \epsilon + p / \rho$ is the total specific enthalpy, $\epsilon$ is the specific internal energy, $\rho$ is proper mass density, $\eta^{\mu \nu}$ is the Minkowski metric, $p$ is pressure, $[u^\mu] = \Gamma(1,\vec{\beta})$ is four-velocity, $\Gamma = (1 - \beta^2)^{-1/2}$ is Lorentz factor, and $\beta$ is the fluid velocity in units of $c$. The signature of the metric is $(- + + +)$. The system of equations is closed by assuming an adiabatic equation of state
$p = (\hat{\gamma} - 1)\rho \epsilon$, where $\hat{\gamma} = 4/3$ is the adiabatic gas index. We work in units where $c = 1$. The point explosions are launched in a 2D axisymmetric spherical-polar mesh in the $(r-\phi)$ plane. All quantities are made dimensionless by combinations of $R_\odot$, $M_\odot$, and $c$. The code used to run the simulations, entitled \texttt{SIMBI}, was written by this study's first author, and is an open-source, second-order, Godunov-type scheme. \texttt{SIMBI} is is parallelized for multi-CPU frameworks enabled by \texttt{OpenMP}, and supports GPU-acceleration on both \texttt{AMD} and \texttt{Nvidia} platforms \citep{Simbi+2023}.
\subsection{Initial Conditions}
Using the \texttt{MESA} \citep{Paxton+2011,Paxton+2013,Paxton+2015,Paxton+2018,Paxton+2019} code, \citet{Farmer+2021} evolved a series of binary stars with a focus on the primary star post helium burning. The authors followed the evolution of said primary stars just before core collapse. The binary system had various initial masses, but constant initial mass ratio $q_0 = M_{2\rm init} / M_{\rm init} = 0.8$, and each star was non-rotating. We use these stars as the progenitors in this study. From knowing the initial mass, $M_{\rm init}$, final mass, $M_1$, and initial mass ratio, $q_0$, we can arrive at the final mass ratio of the system pre-supernova assuming the mass exchange was conservative,
\begin{equation}\label{eq:q}
    q = M_2 / M_1 = M_{\rm init}(1 + q_0) / M_1  - 1.
\end{equation}
This parameter is crucial for constructing the equipotential surfaces of the Roche lobe. The separation of the stars can be calculated by inverting the \citet{Eggleton+1983} formula to arrive at 
\begin{equation}\label{eq:sep}
    d = R_{\rm eq}\left(\frac{0.49q^{2/3}}{0.6q^{2/3} + \ln(1 + q^{1/3})} \right)^{-1},
\end{equation}
where $R_{\rm eq}$ is the volume equivalent radius defined such that the volume of the Roche lobe is $V = 4/3 \pi R_{\rm eq}^3$.
For simplicity, we choose a single primary star from the grid of binary simulations for our entire analysis. This primary had an initial mass of $M_{\rm init} = 11M_\odot$ and pre-supernova mass of $M_1 = 3M_\odot$, which gives a final mass ratio of $q = 5.21$. Its volume equivalent radius is roughly $5.4R_{\odot}$.  

In this analysis, we assume a steady wind environment, 
\begin{equation}\label{eq:ambient}
    \rho_{\rm amb}(r, \phi) = \frac{\dot{M}}{4 \pi v_{\rm wind}R_*^2(r,\phi)}\left(\frac{r}{R_*(r, \phi)} \right)^{-2} \equiv \frac{A}{r^2},
\end{equation}
where $\dot{M}$ is the stellar mass loss rate, $v_{\rm wind}$ is the velocity of the wind, and $R_*$ is the outer radius of the distorted envelope. A common convention for modulating stellar wind magnitudes is the mass-loading parameter $A = A_* \times \unit[5 \times 10^{11}]{g \ cm^{-1}}$, which assumes $\dot{M} = \unit[10^{-5} M_\odot]{yr^{-1}}$ and $v_{\rm wind} = \unit[10^8]{cm \ s^{-1}}$. We fix $A_* = 1$ in our simulations. 

The \texttt{MESA} density profiles are computed in 1D using the volume equivalent radius, $R_{\rm eq}$. Since the equipotential surfaces of the Roche lobe are themselves isobars, we can invoke hydrostatic equilibrium ($- \nabla P / \nabla \Phi = \rho$) to map the density $\rho(R_{\rm eq})$ to the surfaces of constant potential, $\Phi$. We also assume synchronous rotation for the system such that the dimensionless potential has the form \citep{Kopal+1959}:
\begin{eqnarray}\label{eq:potential}
    \Phi(r, \phi)  &=& \frac{1}{r} + \frac{q + 1}{2}r^2 \\
    &+& q\left(\frac{1}{\sqrt{1 - 2r\cos\phi + r^2}} - r\cos\phi \right) \nonumber,
\end{eqnarray}
where we set $\theta = \pi / 2$ so that $r-\phi$ lies in the orbital plane of the binary. The formulae above allow us to transform the 1D density profiles from $\rho(R_{\rm eq})$ to multidimensional profiles $\rho(\Phi)$, conserving the stellar mass in the process. To map the density contours for a Roche lobe filling factor, $\mathcal{F}$, we use the functional form of the modified potential described in \citet{Leahy+2015}:
\begin{equation}
    \Phi_\mathcal{F} = \frac{\Phi_{\rm L1} + q^2/2(1+q)}{\mathcal{F}} - \frac{q^2}{2(1+q)},
\end{equation}
where $\Phi_{\rm L1}$ is the potential at the L1 Lagrange point of the system. 

 We model the supernova explosion as a ``thermal bomb'' in a small region of the grid,
 \begin{equation}\label{eq:pressure}
      p(r) = \frac{3(\hat{\gamma} - 1)E}{4 \pi \delta r} \times \mathcal{H}(r - \delta r),
 \end{equation}
 where $\delta r$ is the radial extent of the overpressure, and $\mathcal{H}(r)$ is the Heaviside step function. Since we are constrained geometrically by the detailed evolution of the binary, we are only afforded the liberty of varying $E/M$ and the Roche lobe filling factor, $\mathcal{F}$. 
Results are presented with $2048$ radial zones 
per decade on a logarithmically-spaced spherical-polar mesh with $\Delta r = r \sin(\pi/2) \Delta \phi$. The radial extent is $r_{\rm min} = 10^{-3}R_\odot$ to $r_{\rm max} = 10R_\odot$ and the azimuthal extent covers half the sphere, $\phi \in [0, \pi]$. This results in a grid of 8192 radial zones by 2795 angular zones.  In other words, the explosion is resolved in 361 radial zones, and the star is resolved within 7644 radial zones. We set $\delta r = 1.5r_{\rm min}$. 
\section{Results} \label{sec:results}
\begin{figure}
\centering
\gridline{\fig{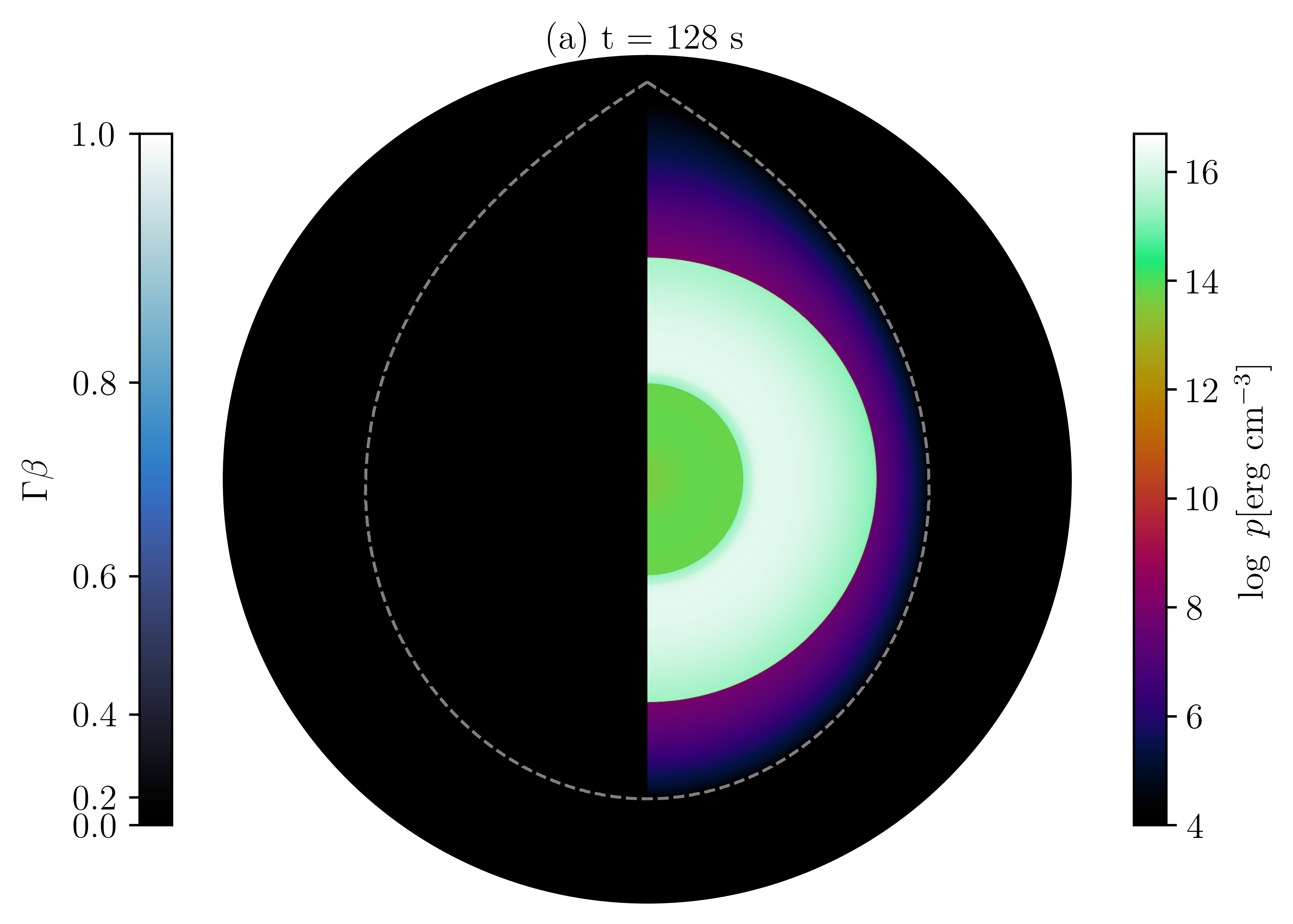}{\columnwidth}{}}\vspace*{-3.0em}
\gridline{\fig{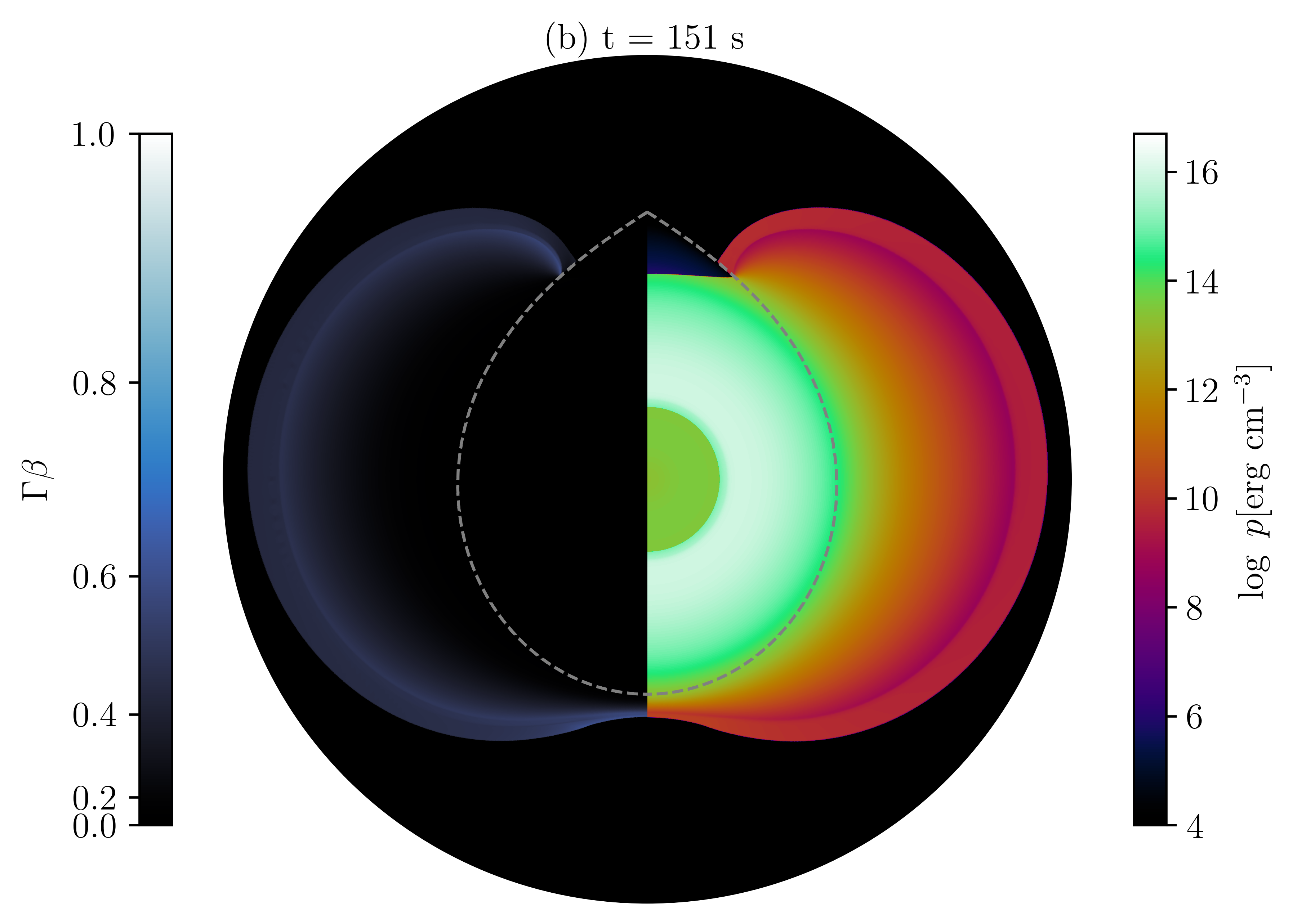}{\columnwidth}{}}\vspace*{-3.0em}
\gridline{\fig{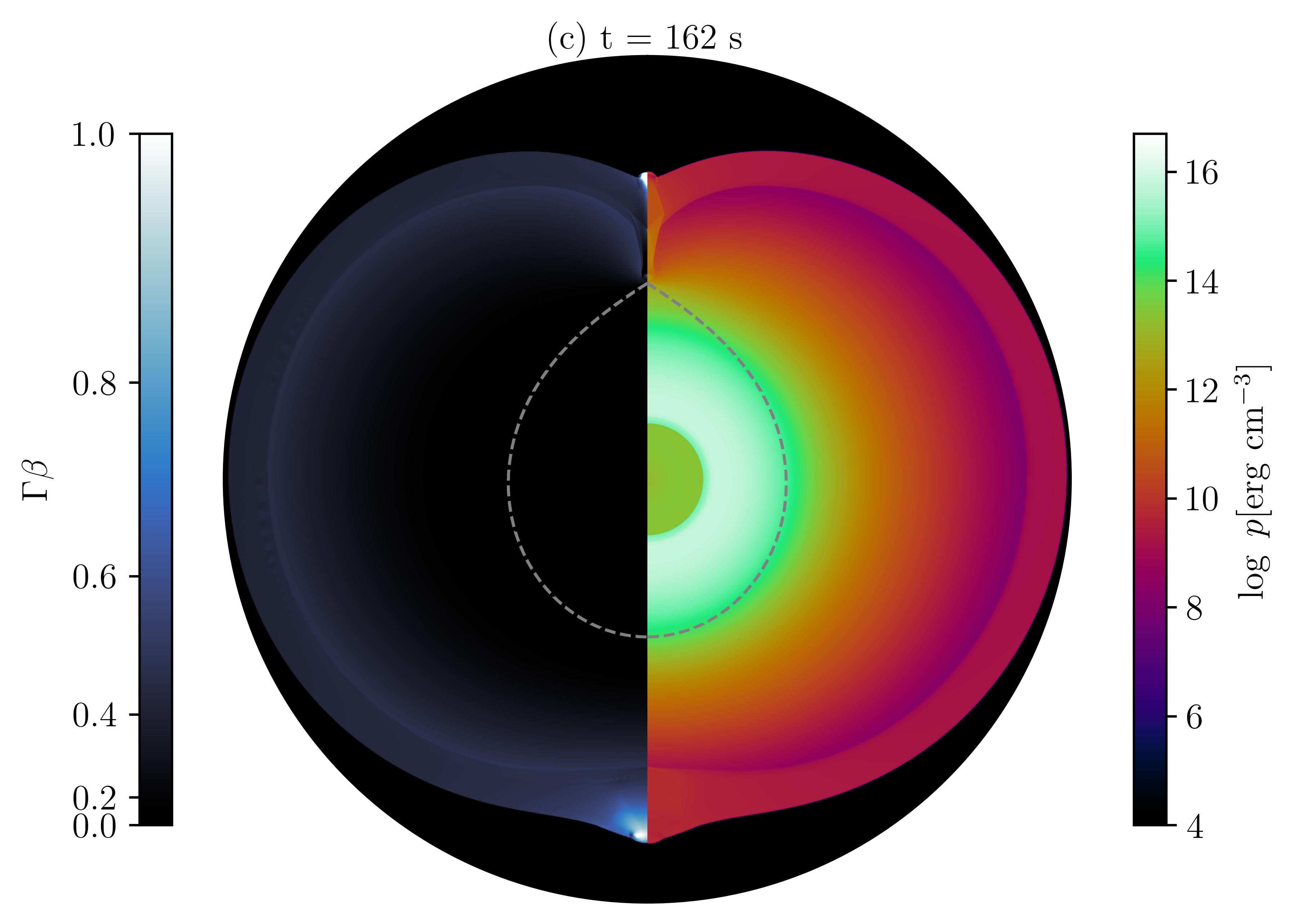}{\columnwidth}{}}\vspace*{-1.0em}
\caption{The dynamical evolution of a \foe{} explosion within a Roche-lobe filling stripped star ($\mathcal{F} = 1$, shown as the grey dashed line with volume-equivalent radius $\sim\unit[3.8 \times 10^{11}]{cm}$). Logarithmic pressure is shown on the right half and four-velocity on the left half. In panel (a), the shock wave evolves quasi-spherically while still inside the stellar mantle. In panel (b), the supernova shock has broken out of the stellar poles first and begins wrapping around the star towards collision points at the Roche-lobe nose and the back of the progenitor. In panel (c), the lobe shocks collide and force material along the center-of-mass axis of the binary, forming two relativistic bullets directed towards and away from the companion with masses $\unit[6 \times 10^{27}]{g}$ and $\unit[5\times10^{25}]{g}$, of order the masses of the Earth and Moon, respectively.}
\label{fig:shock_evol}
\end{figure}
Figure \ref{fig:shock_evol} depicts the shock wave evolution from a \foe{} explosion set off inside the $3M_\odot$ progenitor. The first snapshot of the explosion shows that at first, the supernova shock wave propagates outward quasi-spherically while still interior to the stellar mantle in panel (a). This is because the iso-density contours near the stellar core are quasi-spherical due to gravity, so there is no preferential density gradient in either direction that would initially accelerate and deform the supernova shock \citep{Sakurai+1960,Matzner+Mckee+1999}. However, because of the geometry of the Roche lobe filling star, the shock waves will break out at the stellar poles first, i.e., at $\theta = 0$ and \emph{not} $\phi = 0$, since the ratio of the distance to one of the stellar poles to the distance to the back of the star is $R_z / R_{\rm back} \approx 0.85$. The result of this asynchronous breakout is shown in panel (b) of Figure \ref{fig:shock_evol} where the shock continues its outward expansion, but the oblique breakout causes the shocks to wrap around towards the nose of the Roche lobe as well as the back of the progenitor. The wrap-around effect is caused by the naked blast wave realizing a lateral pressure gradient in the $\hat{\phi}$ direction resulting in the blast wave forming a pattern wave along the stellar surface, pointing towards impending collision points along the center-of-mass axis of the binary \citep[see e.g.,][and references therein]{Matzner+2013,Afsariardchi+2018,Scully+2023}{}{}. Finally, in panel (c) we see the aftermath of the lobe shocks colliding toroidally at a point located along the center-of-mass line of the binary system.\@ The shocked lobes coalesce at this point, forming a hot, pressurized blob of ejecta whose secondary acceleration is caused by the steep pressure gradient directed along the binary center-of-mass axis. This ultimately results in a violent spray of ejecta that form collimated relativistic bullets pointing both away and towards the companion. The front bullet has a mass of $\unit[6 \times 10^{27}]{g}$ and the back bullet has a mass of $\unit[5 \times 10^{25}]{g}$, which are of order the masses of the Earth and Moon, respectively.

\begin{figure}
    \centering
    \includegraphics[width=\columnwidth]{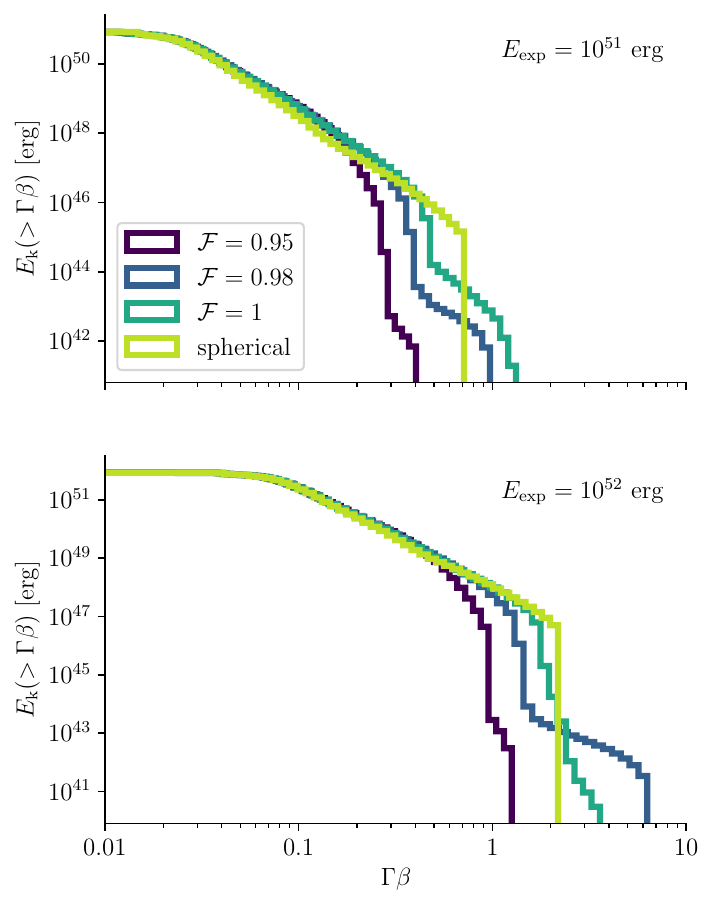}
    \caption{The ballistic cumulative kinetic energy distribution for the \foe{} and \tenfoe{} explosions at $\unit[162]{s}$ and $\unit[58]{s}$, respectively, for all geometries considered. The progenitors with Roche-lobe filling factors $\mathcal{F} \geq 0.98$ extend the high-velocity tail of the ejecta by factors of a few when compared with their spherical counterparts. In general, the $\mathcal{F} = 0.95$ produces the least amount of relativistic material.}
    \label{fig:ek_hist}
\end{figure}

To quantify some of the observable properties of the relativistic ejecta, a useful measure is the cumulative kinetic energy distribution above some four-velocity, $E_k(>\Gamma\beta)$. Figure \ref{fig:ek_hist} plots $E_k(>\Gamma\beta)$ for all geometries and explosion energies considered in this work. There is non-monotonic behavior in the Roche lobe stars where the geometries $\mathcal{F} \geq 0.98$ extend the high-velocity tail beyond the ballistic velocities shown for their spherical counterparts. The $\mathcal{F} = 0.95$ stars, however, produce no material in such high-velocity ejecta and are slower overall than the spherical counterparts. Although the $\mathcal{F} \geq 0.98$ explosions extend the high-velocity tail of the kinetic energy distribution by factors of only a few, the angular distribution of this fast tail shows a clear split in the azimuthal angle. 

\begin{figure}
    \gridline{\fig{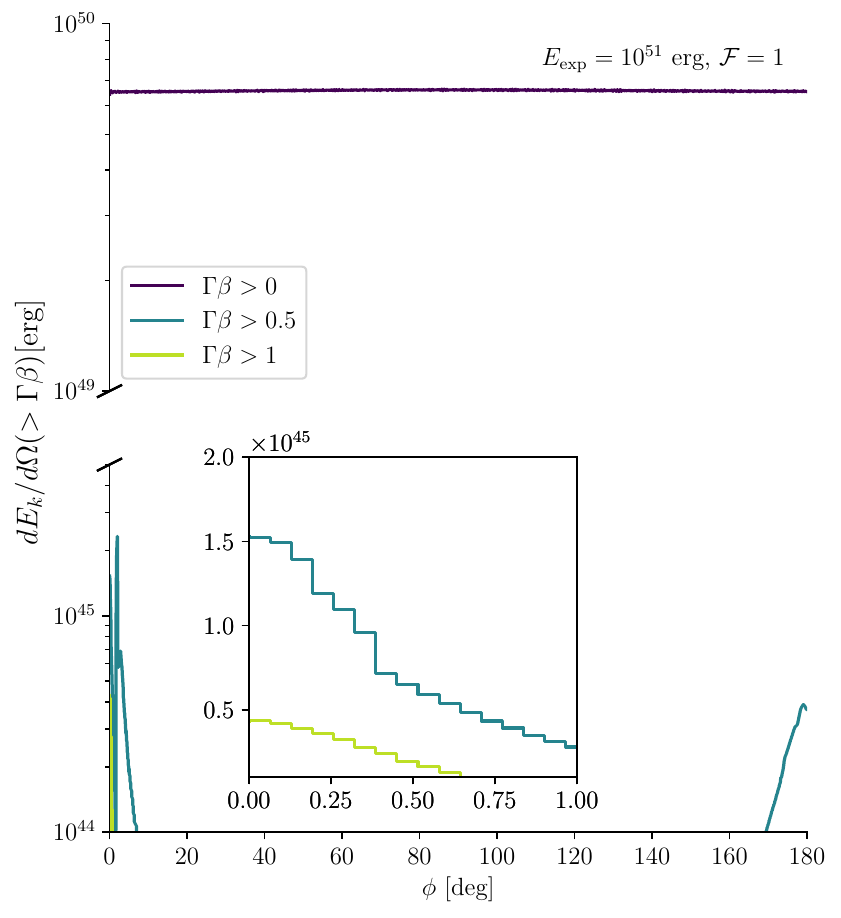}{\columnwidth}{(a)}}
    \gridline{\fig{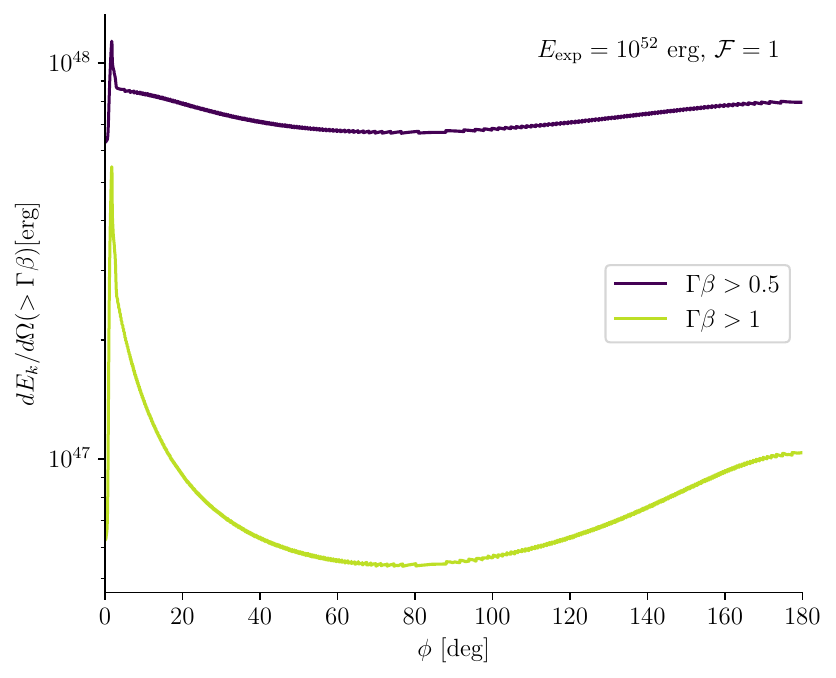}{\columnwidth}{(b)}}
    \caption{\emph{Upper}: The cumulative kinetic energy per unit solid angle for the \foe{} explosion as a function of azimuthal angle, $\phi$, in the star that has filled its Roche lobe, i.e., $\mathcal{F} = 1$. The inset marks the region aimed at the companion where the fastest ejecta is focused, carrying total energy $E_k({>}1) =  \Omega_{\rm bullet} dE_k/d\Omega = \unit[5 \times 10^{41}]{erg}$ towards the companion in a bullet of angular width $\sim 1^\circ$. There is also a counter bullet pointed away from the companion carrying away total energy $\sim \unit[10^{44}]{erg}$ in ejecta with $\Gamma \beta > 0.5$. \emph{Lower}: The cumulative kinetic energy per unit solid angle for the \tenfoe{} explosion where most of the focusing happens towards the companion with $E_k({>}1) =  \Omega_{\rm bullet} dE_k/d\Omega = \unit[2 \times 10^{45}]{erg}$.}
    \label{fig:ek_vs_omega}
\end{figure}

Figure \ref{fig:ek_vs_omega} plots, for the $\mathcal{F} = 1$ progenitors only, the cumulative kinetic energy distribution per unit solid angle for the \foe{} and \tenfoe{} cases at various velocity cutoffs. In panel (a), to demonstrate the global nature of the ejecta, we plot the $\Gamma\beta = 0$ cutoff and can recover the \foe{} explosion energy if multiplied by the full solid angle $4\pi$. However, when we discriminate towards the mildly relativistic ($\Gamma\beta \simeq 0.5$) and relativistic ($\Gamma\beta \simeq 1$) ejecta, we see energy is largely focused along the binary axis for both explosion energies considered. For example, let $dE_k/d\Omega \equiv E_\Omega$. In the \foe{} case we find $E_\Omega ({>} \{0.5, 0.1\})\vert_{\phi=0^\circ} = \{\unit[1.5\times 10^{45}]{erg}, \unit[5 \times 10^{44}]{erg} \}$ and $E_\Omega ({>}0.5) \vert_{\phi=180^\circ} = \unit[4 \times 10^{44}]{erg}$. In the \tenfoe{} case, we find that most of the energy is just focused towards the companion with values $E_\Omega({>}\{0.5, 1\})\vert_{\phi=0^\circ} = \{\unit[10^{48}]{erg}, \unit[5 \times 10^{47}]{erg} \}$. 

The solid angle of the bullets at the different velocity cutoffs is from
\begin{eqnarray}\label{eq:solid_angle}
    \Omega &=& 4\pi \frac{E_{\rm iso}}{E} \nonumber \\
    &=& 4\pi \left(\frac{\int_\Omega \left( dE/d\Omega\right)^2d\Omega}{\int_\Omega (dE/d\Omega) d\Omega} \right)\left(\int_\Omega (dE/d\Omega)d\Omega \right)^{-1},
\end{eqnarray}
where $E_{\rm iso}$ is the isotropic-equivalent energy weighted by the energy per solid angle. With Equation \ref{eq:solid_angle}, we compute $E_k({>}\{0.5, 1\})\vert_{\phi=0^\circ} = \{\unit[2 \times 10^{44}]{erg}, \unit[5 \times 10^{41}]{erg}\}$, and $E_k({>}0.5)\vert_{\phi=180^\circ} = \unit[6 \times 10^{43}]{erg}$ as the total kinetic energies pointed towards and away from the companion for the \foe{} explosion. For the \tenfoe{} explosion, the total energies are $E_k({>}\{0.5, 1\})\vert_{\phi=0^\circ} = \{\unit[4 \times 10^{45}]{erg}, \unit[2 \times 10^{45}]{erg} \}$.

\begin{figure*}\label{fig:money_plot}
\gridline{\fig{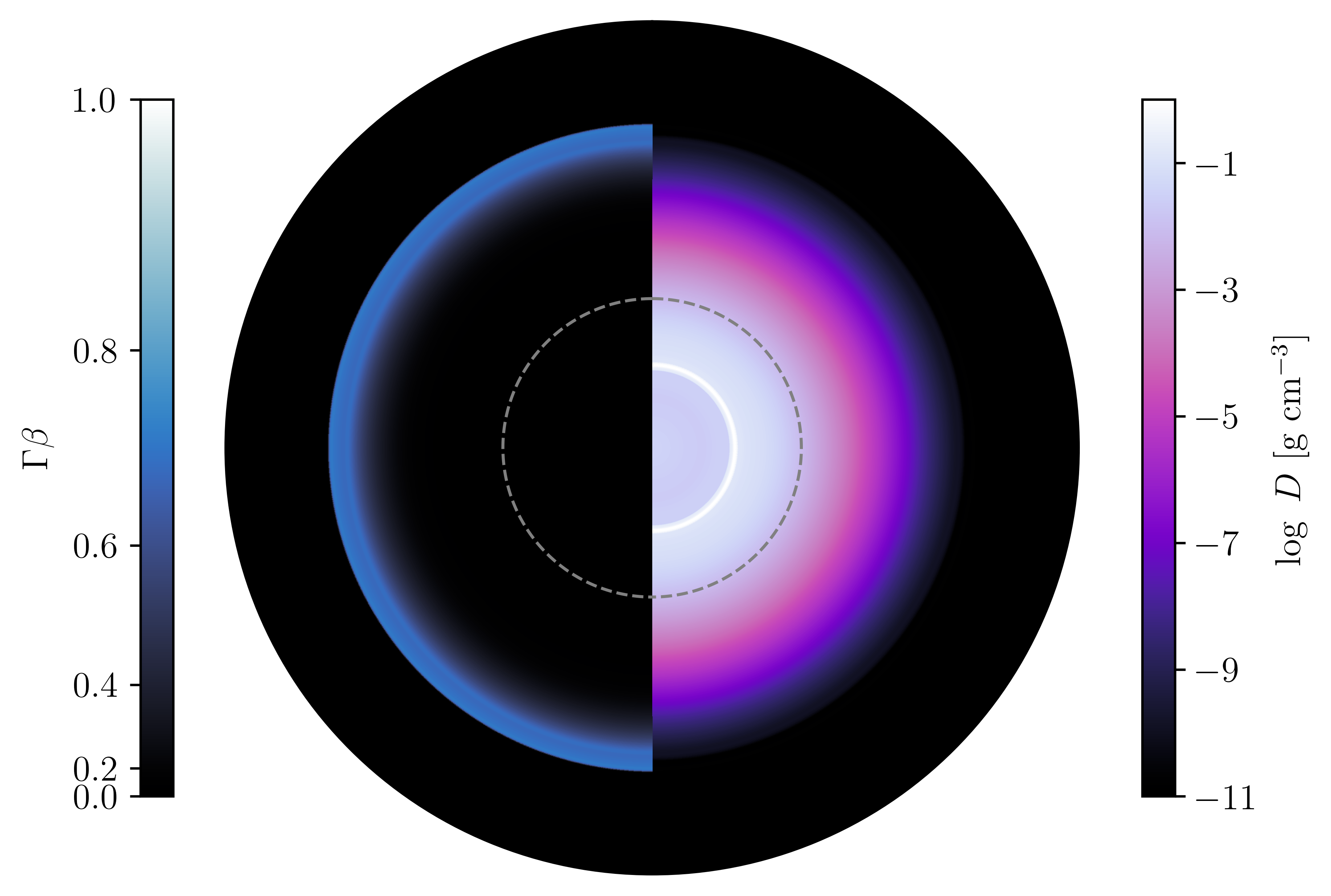}{0.5\textwidth}{(a)}
          \fig{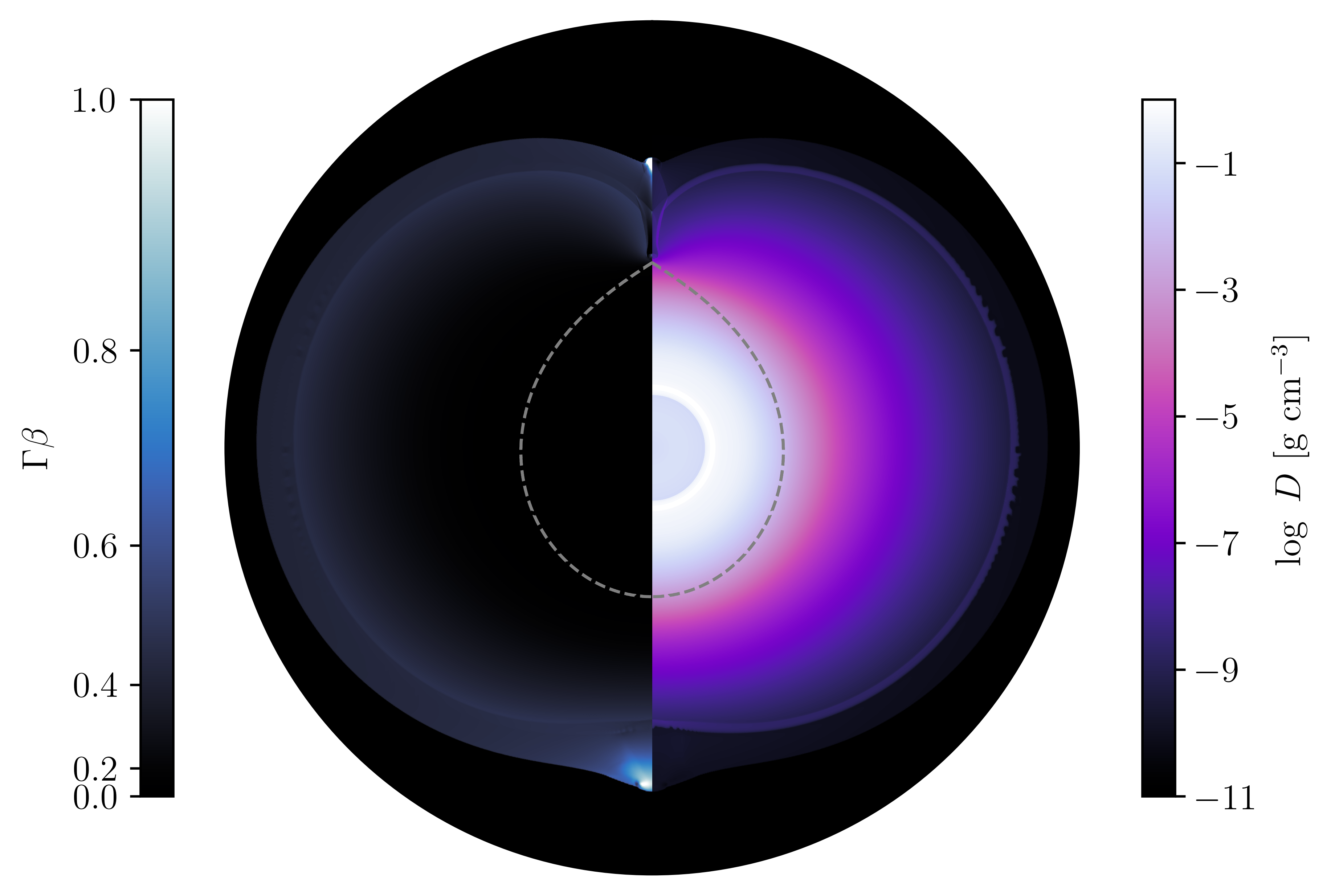}{0.5\textwidth}{(b)}}
\caption{Evolution of the \foe{} supernova a few minutes post explosion in the stripped star (a) and in the same stripped star that has filled its Roche lobe (b). Lab frame density is plotted on the right side and four-velocity is plotted on the left. The grey-dashed curves mark the original outer surface of the $3M_\odot$ progenitor with volume-equivalent radius $\unit[3.8 \times 10^{11}]{cm}$. The outflow velocity for the spherically symmetric star peaks at around $\Gamma\beta = 0.5$ while the fastest ejecta is focused along the binary axis for the Roche lobe filling star and has velocity $\Gamma\beta \gtrsim 1$.}
\end{figure*}

Lastly, Figure \ref{fig:money_plot} shows the separation between \foe{} explosions in the canonical spherical progenitor and a star that has filled its Roche lobe. At freeze out of the acceleration in both cases, the Roche lobe filling star produces trans-relativistic ejecta that gets focused along the binary separation axis while there is no such fast ejecta in the perfectly spherical supernova explosion. This distinctive asymmetry in the velocity and density profiles might lend itself to various hints at the supernova signatures observed and are further discussed in the next sections. We remind the reader that these results are idealized to 2D axisymmetry, which may enhance the effect of the shock converging at the nose and backside of the Roche-lobe. Exploration of the 3D nature of this problem will be needed and we plan to undertake this in a future study.

%
%
\section{Astrophysical Implications} \label{sec:discussion}


\subsection{A relativistic bullet aimed at secondary and implications for shock breakout\label{sec:dis_4_1}}
The late-stage morphology of explosion in the Roche-lobe distorted star implies that the companion lies along the path of one of the  relativistic bullets.  For the companion star to dodge the relativistic bullet, its sectorial orbit would need to sweep an angle greater than the sum of the solid angle of the companion with that of the bullet which leads to the companion radial constraint: 
\begin{equation}\label{eq:rconstraint}
    \frac{R_{\rm C}}{d} < \tan\left(\Delta\phi_{\rm orb} - \Delta\phi_{\rm bullet}\right),
\end{equation}
where $R_{\rm C}$ is the radius of the companion, $d$ is the binary separation, $\Delta\phi_{\rm orb} \equiv \tan^{-1}(V_{\rm orb}/\beta_{\rm bullet} c)$ is the sectorial width of the companion orbit in the time it takes the relativistic bullet to reach the companion's original position, $V_{\rm orb} \approx (GM/d)^{1/2}$ is the orbital velocity of the companion, $\beta_{\rm bullet}$ is the velocity of the ballistic bullet, and $\Delta\phi_{\rm bullet}$ is the half-opening angle of the bullet. With the bullet moving at the speed of light and having $\Delta\phi_{\rm bullet} \simeq 0.01$, Inequality \ref{eq:rconstraint} requires $V_{\rm orb} > 0.01c$ in order for the argument in the tangent function to be non-negative.
Since typical orbital velocities in binary systems don not exceed a few hundred kilometers per second, this means that the bullet will hit any companion regardless of its type whether it is a neutron star (NS), a black hole (BH), or a main sequence (MS) star. We reserve discussion of compact companions for later sections since their predicted rates are lower than those of MS companions \citep[see e.g.,][]{Zapartas+2017,Renzo+2019}.

While the energy budget of the directed bullet is not enough to unbind typical envelopes for massive companion stars, the deposition of energy could potentially affect the companion's evolution through induced mixing effects, seismic activity, or the collision could release a burst of radiation visible from the shadow cone \citep{Kasen+2010,Ogta+2021}. Furthermore, the companion star will be shock heated, expand, and radiate, which will produce late-time radiation signatures at the supernova location \citep[see e.g.,][for up-to-date observations of stripped-envelope SNe with companion detections]{Maund+2004,Folatelli+2014,Maund+2016,Ryder+2018,Maund+2019,Sun+2020}{}{}, but since the energy imparted by the bullet will be small, the companion's inflation timescale will be shorter lived than values expected for spherically symmetric SN hitting their companions. We also note that if the explosion energy is large enough, the ejecta-ejecta collision along the binary axis could produce a flash of photons before the SN light is seen as discussed in \citet{Scully+2023}, and the combined energies of the bullets in the \tenfoe{} explosion simulated in this work are  $\sim\unit[10^{46}]{erg}$, roughly the released energy of XT 080901 \citep{Soderberg+2008}. Moreover, the stripped-envelope progenitors studied in this work are possible candidates for rapidly evolving SNe now better known as a subclass of fast blue optical transients \citep[FBOTs;][]{Drout+2014,Kleiser+2014,Kleiser+2018+a,Kleiser+2018+b}, though the luminosity peak will shift since the breakout flash is partially obscured by the obliquity of the breakout \citep{Matzner+2013}. All of these together --- the ejecta-ejecta flash, short shock cooling timescale of the companion, and non-standard light curve morphology --- might provide hints towards classification of asymmetric SNe in binaries versus their spherically symmetric counterparts.   




\subsection{Polarization signatures \label{sec:dis_4_2}}
In principle, polarization measurements can be made if there is asymmetry in the supernova progenitor itself or if the explosion engine was asymmetric at the outset \citep{Hoeflich+1991,  Yamada+1991, Hoeflich+1996}. In the case for the Roche lobe bullets discussed in this work, the asymmetric outflow will likely give rise to measurable non-zero net polarization, assuming that it is dominated by Thompson scattering on free electrons. To get a sense of the lifetime of the asymmetry in the flow, we estimate the deceleration of the backwards bullet based on results from our simulations and assumptions about the stellar environment. The front bullet will be quickly destroyed by interactions with the companion, but the backwards bullet will decelerate after a time $\left(\frac{M_{\rm bullet}}{4\pi A} \right) / \beta c$, which, when scaled for a few reference parameters from the \foe{} explosion, gives
\begin{equation}\label{eq:tdec}
   t_{\rm dec}  = \unit[12 \left(\frac{M_{\rm bullet}}{\unit[5 \times 10^{25}]{g}} \right) \left(\frac{A_*}{10^{-3}} \right)^{-1}\left(\frac{\beta}{0.5} \right)^{-1}]{days}
\end{equation}
and we have chosen $A_* = 10^{-3}$ such that asymmetry of the outflow lasts $\sim$ week which is roughly the earliest cadence at which we can currently observe SNe post explosion \citep[see][and references therin]{Wang+Wheeler+2008,Patat+2017}. Equation \ref{eq:tdec} is two-pronged. That is to say, the stellar environment surrounding the Roche lobe explosion would have to be quite rarefied or the mass of the bullet (and therefore the explosion energy) would have to be larger in order to measure polarization signatures for these types of explosions. At least one of these conditions can be satisfied based on: (i) the large upper limits placed on the explosion energies of core-collapse events; (ii) the stellar environment surrounding stripped-envelope stars being uncertain. Furthermore, if the asymmetry lasts much longer than we estimate, it is possible that bullets in binaries could produce core-collapse SN remnants with ``ears'' \citep{Grichener+2017}.

\subsection{Asymmetric engines \label{sec:dis_4_3}}
Although this work is in the context of spherically symmetric explosions in aspherical progenitors, there lies a plethora of observational evidence that the resultant NS star formed from core-collapse can have a natal kick from an asymmetric explosion mechanism \citep[see e.g.,][]{Hobbs+2005,Scheck+2006, Wongwathanarat+2012,Janka+2017}. An asymmetric explosion in an asymmetric progenitor will likely greatly change the distribution of energy over solid angle shown in Figure \ref{fig:ek_vs_omega} since the resultant bullet velocities strongly depend on the time and the angle at which they collide along the binary axis. The advantage of the point explosion is the exterior lobe shocks colliding simultaneously with the interior SN shock that breaks out at the Roche lobe nose. This advantage of simultaneous collisions depends on the explosion energy and $\mathcal{F}$ as shown in Figure \ref{fig:ek_hist}. When the explosion itself is asymmetric, the dynamics of the collision outside of the Roche lobe filling star will change in a non-trivial way and should be studied in a future work. 

\subsection{Fast Radio Bursts}\label{sec:dis_4_4}
In the scenario in which the binary companion is a neutron star (NS), the relativistic bullet aimed toward the companion will encounter the strong magnetic field of the NS magnetosphere. In this case, as the plasma bullet hits the magnetic field and is decelerated, some of its kinetic energy can be emitted in the form of radio waves. The mildly relativistic bullet has energy $\sim\unit[10^{44}]{erg}$ and initial radius of $\unit[\sim 3 \times 10^{7}]{cm}$. Assuming the bullet is stopped almost instantaneously by the strong NS magnetic field, then the time for the reverse shock to traverse the bullet is roughly $\unit[10^{-3}]{s}$, of order the duration of a fast radio burst \citep[FRBs;][]{Lorimer+2007}. We use the plasma frequency, $\nu$, as an estimate for the radio wave frequency,
\begin{eqnarray}\label{eq: plasma_freq}
    \nu = \frac{\omega_{\rm plasma}}{2\pi} &=& \left(\frac{n_e \Omega_{\rm B}e^2}{\pi m_e \Omega_{\rm bullet}} \right)^{1/2} \nonumber \\
    &=& c \left(\frac{r_en_e}{\pi} \right)^{1/2}\frac{(R_{\rm SO} / d)}{\theta_0},
\end{eqnarray}
where $n_e = \rho / m_p$ is the electron number density in the bullet, $m_p$ is the proton mass, $e$ is the elementary charge, $m_e$ the electron mass, $r_e$ is the classical electron radius, and we've accounted for the ratio of solid angles of the bullet with that of the magnetosphere of the NS where $R_{\rm SO}$ is the stand-off radius with value:
\begin{eqnarray}\label{eq:rstandoff}
    R_{\rm SO} &\sim& R_{\rm NS} \left(\frac{B_d^2}{8\pi \rho u^2c^2} \right)^{1/6} \nonumber \\ 
    &\approx& 3 R_{\rm NS}B_{d, 12}^{1/3}\rho_{-1}^{-1/6}u_{1}^{-1/3},
\end{eqnarray}
where $B_d$ is the surface dipole field of the NS and we follow the convention that quantities $Q_x = Q / 10^x$ with cgs units implied. From this, we find
\begin{equation}
    \nu \simeq \unit[3 \times 10^{12}B_{d,12}^{1/3}\rho_{-1}^{1/3}d_{12}^{-1}u_1^{-1/3}\theta_{0,-2}^{-1}]{Hz}.
\end{equation}
This implies that the bullet meets the energy, duration, and frequency requirement for an FRB if it crashes into an NS magnetosphere, assuming an efficiency $10^{-5}-10^{-4}$ for converting the kinetic energy into the isotropic radio emission of $\unit[10^{39}]{erg}-\unit[10^{40}]{erg}$ quoted for many FRBs \citep[see][for a detailed review of the current state of FRB science]{Cordes+2019}. Based on recently calculated rates of SNe in the universe \citep[$\sim 1$ s$^{-1}$][]{Li+2011} and FRB rates \citep[$\sim 0.1$ s$^{-1}$][]{Thornton+2013} being in rough agreement with the rate of NS companions existing in close binaries with massive stars \citep{Zapartas+2017,Renzo+2019}, we predict that supernova explosions in close binaries with a magnetized NS companion are a viable source for non-repeating FRBs via this mechanism. In this scenario, the FRB would be followed by the rising light curve of a stripped-envelope supernova (SN Type Ib/c) several days later.     

\subsection{Relevance for the formation of Double NS and GW sources \label{sec:dis_4_5}}

Explosions of Roche-Lobe distorted stars are expected in the main binary formation scenarios for double compact objects. This is especially true in the main formation pathway for double neutron stars \citep[e.g.][]{Dewi2003,Tauris+2017}, where both progenitors may be significantly distorted. Our findings show potentially observable signatures that could be used to search for systems evolving through this formation channel, see for example \citet[][]{De2018}. 

The first explosion is typically coming from the star that was once the most massive star in the system. Depending on the separation and the amount of swelling of the progenitor, this may already be an example of a Roche-distorted star with an main sequence companion. Observable consequences may result from ejecta hitting the companion, which is most likely still a main sequence star in this phase (Sect.~\ref{sec:dis_4_1}) and the aspherical outflows which may give rise to polarization signatures (Sect.~\ref{sec:dis_4_2}).  

Most systems unbind during the first explosion, due the supernova kick resulting from asymmetries in the explosion \citep[e.g.][]{Hobbs2005}.  Although we show in this work that additional asymmetries may result from the deformation of the star, we find that these will not play any significant role for the momentum imparted onto the system and the newly born neutron star. 

Systems that remain bound can evolve further as a binary system. The second star now evolves to fill its Roche-lobe. It first loses its hydrogen-rich envelope in a highly non conservative event during which the orbit shrinks. The result is a compact helium star orbiting a neutron star in a tight orbit.  The helium star is expected to swell once its start burning helium in a shell around its core (often referred to as case BB mass transfer). It will fill the Roche lobe anew now feeding gas to the neutron star, a stage that may be observable as an X-ray binary. Eventually the helium star will complete the advanced burning stages it will explode too, likely while still filling its Roche-lobe. At this stage we expect a heavily Roche-lobe distorted star with a neutron star companion, possibly with the observational consequence of a fast radio burst  (Sect.~\ref{sec:dis_4_4}) accompanied by a supernova.

\section{Summary and Conclusions}\label{sec:summary}
This work has demonstrated the following: (1) point-like explosions in pear-shaped stripped-envelope stars  cause shocks to converge to a point, producing relativistic bullets pointed away and towards their companions even for the typically non-relativistic \foe{} explosions; (2) such explosions only impart $\sim 10^{44} - 10^{45}$ erg onto their companions, drastically reducing the shock cooling timescales for main sequence companions assumed for spherical SN in binaries; (3) these explosions can obscure the breakout flash and therefore the luminosity peak as seen for many rapidly evolving SNe; (4) point-like explosions in pear-shaped stripped-envelope stars might be candidates for Fast Radio Bursts associated with a core-collape SNe if the relativistic bullet interacts with the magnetic field of a neutron star companion. In short, as the vampire companion star depletes the primary of its envelope, at the end of the primary's life it responds by blasting the vampire star with a relativistic bullet.

\begin{acknowledgments}
This work was supported in part by NASA ATP grant 80NSSC22K0822. MD and SdM thank R\"udiger Pakmor, Mathieu Renzo, Shazrene Mohamed, Nando Patat, Rob Farmer, Stephen Justham, Eva Laplace, and Norbert Langer for useful discussions and comments during the MPA-Kavli summer school 2023, and MD acknowledges useful discussions with Andrei Gruzinov and support from LSSTC Data Science Fellowship Program, which is funded by LSSTC, NSF Cybertraining Grant \#1829740, the Brinson Foundation, and the Moore Foundation. 
\end{acknowledgments}

%

\vspace{5mm}


\software{\texttt{CMasher} \citep{CMasher}, \texttt{SIMBI} \citep{Simbi+2023}, \texttt{MESA} \citep{Paxton+2011,Paxton+2013,Paxton+2015,Paxton+2018,Paxton+2019}, \texttt{astropy} \citep{astropy+2013}}




\bibliography{refs}{}
\bibliographystyle{aasjournal}



\end{document}